# QoS and Coverage Aware Dynamic High Density Vehicle Platooning (HDVP)

Yinan Qi, Tomasz Mach

Samsung Electronics R&D Institute UK, Staines, Middlesex TW18 4QE, UK

yinan.qi and tomasz.mach@samsung.com

*Abstract*—In a self-driving environment, vehicles communicate with each other to create a closely spaced multiple vehicle strings on a highway, i.e., high-density vehicle platooning (HDVP). In this paper, we address the Cellular Vehicle to Everything (C-V2X) quality of service (QoS) and radio coverage issues for HDVP and propose a dynamic platooning mechanism taking into account the change of coverage condition, the road capacity, medium access control (MAC) and spectrum reuse while at the same time guaranteeing the stringent QoS requirements in terms of latency and reliability.

## I. Introduction

Vehicle platoons (a.k.a. road trains) has been researched and studied in different industry organizations [1] from a communication perspective as one of the connected vehicle applications benefiting transport and logistics. In the platoon, first vehicle (leader) is driven manually or automatically and following vehicles are controlled by using Vehicle to Vehicle communication. Platoon control could be either centralized or distributed with various degree of control assigned to the leader. Low latency reliable connectivity is an enabler of the platoon application to ensure stable string of vehicles with reduced time headway between them. Platooning pilots are currently being tested in projects mostly in Europe e.g. 'ETPC 2016' [2], 'Sweden4platooning' [3], 'Ensemble' [4].

High Density Vehicle Platooning (HDVP) can further reduce the distance between vehicles down to 1 meter and thus has multiple benefits, such as fuel saving, accident prevention, better traffic efficiency and road infrastructure utilization, reduced costs and $CO_2$ emissions, increased productivity and lower driver workload etc. [5]. Since performance of on-board sensors such as radar or camera only may not be able to fulfill the safety requirements with shorter inter-vehicle distances, platoon vehicles need to continuously exchange their positioning and dynamic kinematic state information in a real time to provide automated lateral and longitudinal control. This will allow following vehicles to use accelerator or brake controls, in order to adjust the the target distance, which requires cooperation among all participating vehicles in order to form, maintain and deactivate the platoon in case of dynamic road situations.

The main challenge for HDVP is how to guarantee platoon stability and safety by providing required communication reliability and latency. As identified in [5] to [7], a maximum C-V2X wireless network end-to-end delay of 5 ms and transmission reliability of 99.999% should be guaranteed to deliver the required application safety performance. It should be emphasized that these requirements need to be achieved subject to radio resource availability, e.g. spectrum. Spectrum plays an important role in realizing the 5G potential in verticals such as vehicular communications. Frequency bands below 6 GHz are most likely to be used because they provide better coverage in comparison to higher frequencies. However spectrum below 6 GHz could be scarce because it is often very crowded by use of other wireless applications. In order to improve communication system capacity and maintain the highly reliable and real time information exchange within a platoon and support intra- or inter-platoon coordination and signalling via V2V or V2X communications, efficient medium access control (MAC) mechanism and spectrum reuse approaches need to be developed. The European Commission funded ONE5G project [8], which investigates "5G Advanced" evolved air interface solution and aims at tuning 5G to meet requirements in multi-service and multi-environment situations. As part of the project, new vertical services, e.g., V2X URLLC service, have been studied.

In this paper, we address the quality of service (QoS) and radio coverage issues for HDVP and propose a dynamic platooning mechanism to guarantee the stringent latency and reliability requirements for V2X URLLC service. In section II, background information and system model are introduced. We analyze the QoS requirements, MAC efficiency and spectrum reuse in section III. Based on the analysis, a dynamic platooning mechanism is also proposed in section III. Evaluation results are presented in section IV and the final section concludes the paper.

## II. System Model

In vehicular communications for HDVP, various transmission modes are supported to enable two major automotive use cases:

- Autonomous device-to-device (D2D) communications between vehicles, i.e., V2V communications, to exchange real-time information between vehicles traveling at fast speeds, in high-density traffic, and even outside of mobile network coverage areas. In this transmission mode, contention-based MAC protocol is employed and packet collision could happen since there is no coordination between vehicles;
- Network scheduled D2D where a base station (BS), e.g., LTE Macro cell or 5G New Radio (NR) gNB, acts as a central controller/coordinator and schedules the exact resources used by a vehicle to transmit direct data. Different from direct D2D, packet collision can be partially or fully avoided depending on the level of BS coordination. Contention-free MAC protocol can be employed to fully avoid packet collision. However, this transmission mode and contention-free MAC protocol can only be used inside mobile network coverage areas.

These two modes of MAC are specified for D2D transmission in the 3GPP standardization Release 12 to 14 [9]-[10]. The autonomous mode, i.e., ad hoc mode, is also considered in IEEE 802.11p [11], where carrier-sense multiple access (CSMA) is the specified MAC scheme for the first generation of vehicular ad hoc networks (VANETs).

In cellular-based D2D systems, depending on coverage condition, D2D communication can be classified into the following coverage scenarios shown in Fig. 1: 1) in-coverage scenario where the devices are in coverage of the BS; and 2) out-of-coverage scenario where the devices are out of coverage. In Fig. 1 (a) network scheduled contention-free MAC protocols can be employed with coordination from the BS inside the coverage area. One the contrary, when vehicles are outside the coverage are as shown in Fig. 1 (b), autonomous contention-based MAC protocols can be employed.

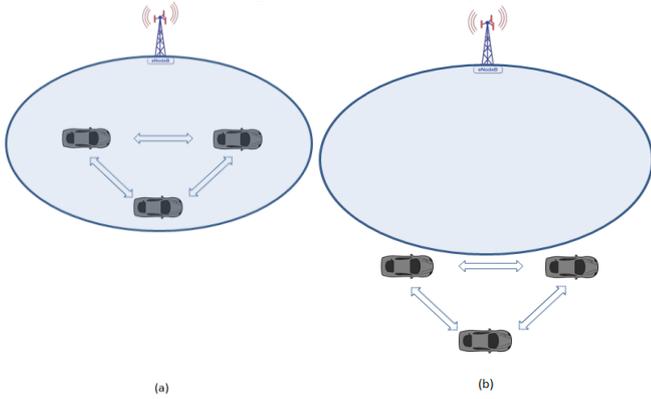

Figure 1 MAC protocols

We assume that all the vehicles in a platoon share the same radio resources. With a given frequency band, different MAC protocols will accommodate different numbers of vehicles and result in different levels of QoS. For example, if network scheduled contention-free MAC protocols are employed, more vehicles can be accommodated in a platoon without violating the reliability requirement, i.e., a larger platoon size can be supported as shown in Fig. 2. If network coordination becomes unavailable, e.g., due to coverage issue, autonomous contention-based MAC protocols need to be employed as shown in Fig. 3, where packet collision might happen. In this paper, subject to stringent QoS requirements and changing coverage conditions, a dynamic platooning mechanism is designed to optimize the road capacity.

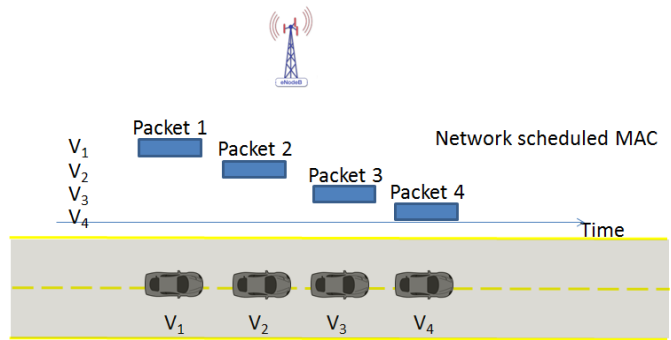

Figure 2 Network scheduled MAC

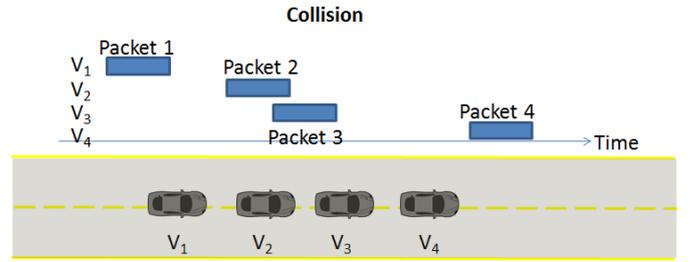

Figure 3 Autonomous contention-based MAC with collision

## III. QoS AND EFFICIENCY ANALYSIS

According to [12], road capacity may be increased by the use of tightly spaced intra-platoon vehicles. The formulation to determine road capacity is as follows:

$$C = v \frac{n}{ns + (n-1)d + D} \quad (1)$$

where $d$ represents the intra-platoon spacing, $D$ is the inter-platoon spacing, $s$ is the vehicle length, $v$ is the steady-state speed, and $n$ is the number of cars in each platoon. As can be seen from (1), road capacity is a monotonically increasing function of platoon size $n$ if we keep intra- and inter platoon distance constant. However, with increased $n$ the inter-vehicle coordination becomes more complex since more vehicles are involved and safety issue may become more critical. In practice, there may be also some regulatory road infrastructure specific limitations e.g. due to increased risk of blocking junctions and interfering with motorway entry or exit for non-platoon vehicles. In this regard, an upper limit on platoon size, denoted as $N_c$, should be imposed. More importantly, the platoon size is also affected by available resources, e.g., spectrum bands, and the stringent QoS requirements of communication between vehicles, such as latency and reliability, which then highly depend on the employed MAC protocol and spectrum reuse approach.

According to [13], the requirement bandwidth to support communications between $N_v$ vehicles can be expressed as

$$B = \frac{L_{pkt} R_{gen} N_v}{S_{mcs} \eta_{MAC}}, \quad (2)$$

where $L_{pkt}$ is the size of the transmitted data packet, $R_{gen}$ is the packet generation rate, $S_{mcs}$ is the spectrum efficiency and $\eta_{MAC}$ is defined as medium access efficiency, describing the level of access coordination. $\eta_{MAC}$ ranges between 0 and 1 and $\eta_{MAC}=1$ means perfectly coordinated medium access and $\eta_{MAC}=0$ means uncoordinated medium access. The level of coordination of different MAC protocols varies and their medium access efficiency can be given as

$$\eta_{STF} > \eta_{Res} > \eta_{CSMA} > \eta_{ALOHA}, \quad (3)$$

where $\eta_{STF}$, $\eta_{Res}$, $\eta_{CSMA}$ and $\eta_{ALOHA}$ are medium access efficiency for static TDM/FDM, reservation-based, CSMA and ALOHA protocols [13].

Eq. (2) can be applied to a single platoon or multiple platoons sharing the same spectrum band. However, for the latter case the MAC protocols should be applied to multiple platoons, causing extra complexity and larger latency. Therefore, it would be more efficient to consider some inter-platoon spectrum sharing mechanism in the multiple platoon case. Taking this into consideration, eq. (2) can be further developed as

$$B = \frac{L_{pkt}R_{gen}N_v}{S_{mcs}\eta_{MAC}\eta_B}, \quad (4)$$

where $\eta_B$ is defined as spectrum reuse efficiency to describing the level of spectrum sharing/management as analyzed in later section III.B. Following (4), the maximum number of vehicles in one platoon can be expressed as

$$N_v = \frac{BS_{mcs}\eta_{MAC}\eta_B}{L_{pkt}R_{gen}}. \quad (5)$$

Eq. (5) clearly indicates that $N_v$ depends on bandwidth, traffic model, spectrum efficiency and MAC and spectrum reuse efficiency, which leads to an important conclusion that the platoon size should be adjusted when these factors change. In the following analysis, we focus on MAC and spectrum reuse.

*A. MAC efficiency*

In this section, we assume the spectrum is occupied by a single platoon, e.g., $\eta_B=1$. As mentioned previously, the level of coordination as well as the MAC efficiency increases from contention-based to contention-free MAC protocols. Their performance also varies depending on service requirements in terms of latency and reliability. Here we consider two typical MAC protocols: slotted ALOHA and Reservation-based MAC. It should be noted that the same analysis can be extended to other MAC protocols, such as CSMA, CSMA/CD, etc.

Slotted ALOHA is a distributed MAC protocol and does not need a central controller. The MAC efficiency can be defined as

$$\eta_{MAC} = \frac{N_v}{N_s}, \quad (6)$$

where $N_s$ is the number of slots to serve all users and can be expressed as

$$N_s = \frac{S_{mcs}B}{L_{pkt}R_{gen}}. \quad (7)$$

We consider two QoS requirements: reliability and latency. In slotted ALOHA, each vehicle just transmits when there is a packet to send and there is a chance that two vehicles transmit at the same time and collision happens. For reliability calculation, we only take collision into consideration for simplicity, i.e., transmission is assumed to be successful as long as no collision happens. The collision probability $P_c$ can be calculated as below and it should be constrained by a given reliability $P_{target}$ in order to fulfill the QoS requirements,

$$P_c = \sum_{N_v}\left[1-\left(\frac{N_s-1}{N_s}\right)^{n-1}\right]\Pr(N_v=n) \le P_{target}. \quad (8)$$

Based on (6)-(8), $\eta_{MAC}$ can be calculated and the maximum vehicles to be supported can be obtained. Latency might be another constraint but reliability is a more critical constraint and imposes more stringent requirement in slotted ALOHA.

We also consider a reservation-based MAC protocol, where each vehicle needs to transmit a preamble to a central controller to reserve medium resources before transmitting a data packet. With reservation, all packets can be put in a queue as long as $N_v \le N_s$, and therefore no collision happens, i.e., reliability requirement can always be met. However, it will increase the latency and the average latency can be given in (9) and should be smaller than the target latency $T_{target}$ as

$$T = \sum_{N_v}\left[1-\left(\frac{N_s-1}{N_s}\right)^{n-1}\right]\Pr(N_v=n)\frac{nL_{pkt}R_{gen}}{S_{mcs}B} \le T_{target}. \quad (9)$$

With (5)-(6) and (9), we can get the maximum platoon size.

One important applicable scenario considered here is the in and out coverage transition. A BS is needed in the reservation-based approach to act as a central controller to coordinate medium access of all vehicles in a platoon. However, this can only be done when the platoon is in the coverage area. Under such circumstances, the platoon size is given by

$$N_{v,in} = \min\{N_{v,res}, N_c\}, \quad (10)$$

where $N_{v,res}$ is the number of vehicles supported by reservation-based MAC and is calculated based on (5)-(6) and (9) amd $N_c$ is an upper limit on platoon size due to safety regulations. Once the platoon is out of the coverage area, the platoon size can be given as

$$N_{v,out} = \min\{N_{v,A}, N_c\}, \quad (11)$$

where $N_{v,A}$ is the number of vehicles supported by ALOHA protocol and is calculated based on (5)-(8). Based on our evaluation results in section IV, $N_{v,in}$ is significantly larger than $N_{v,out}$. It means when the platoon moves out of the coverage area, it needs to be divided into smaller platoons to guarantee that same QoS requirements can be maintained.

*B. Spectrum reuse efficiency*

We consider both single channel and sub-channelization. For single channel, the entire frequency band $B$ is shared by all the platoons. The $N_v$ obtained from the previous section is no longer the platoon size but actually the summation of the sizes for all platoon using the same single channel. In this case the spectrum reuse efficiency is 1. However, it will cause inter-platoon interference so that inter-platoon collision could happen. This might not be a problem when the platoon is in coverage area where a BS can coordinate MAC but could cause some problem in out-coverage area as shown in Fig. 4.

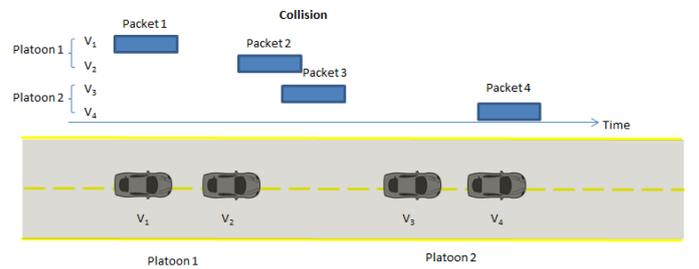

Figure 4 Platoon splitting

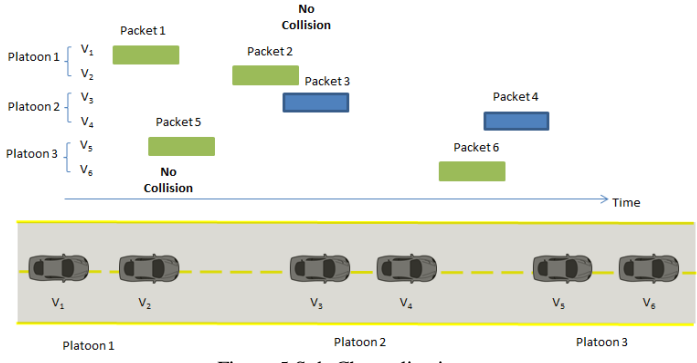

Figure 5 Sub-Channelization

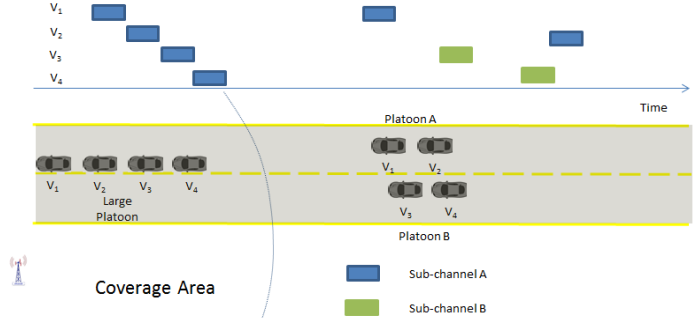

Figure 6 Platoon splitting

In Fig. 4, the supported platoon size becomes smaller when the platoon moves out of the coverage area and platoon splitting happens. However, even though the platoon is divided into two geographically separated smaller platoons, they still share the same spectrum and the collision could still happen as long as they are in the transmission range of each other. The only solution is to let platoon 1 slow down until platoon 2 moves out of platoon 1's transmission range so that there won't be any inter-platoon collision. This could reduce traffic efficiency and potentially cause congestion.

If the entire frequency band $B$ is divided into $N_b$ sub-channels, each with bandwidth $B/N_b$, each sub-channel can either be owned by a single platoon or shared by a few platoons. For the first case, since the sub-channels used by different platoons are orthogonal there is not any inter-platoon interference. The $N_v$ obtained from the previous section is the platoon size. For the latter case, there is inter-platoon interference so that inter-platoon collision may happen and $N_v$ is actually the summation of the sizes for all platoon using the same sub-channel. For both cases the spectrum reuse efficiency is $1/N_b$. The platoons or groups of platoons occupying the same sub-channel can be separated geographically as shown in Fig. 5. In Fig. 5, platoon 1 and platoon 3 use the same sub-channel and but since they are geographically separated, i.e., not in the transmission range of each other, there is no inter-platoon interference.

With sub-channelization, when a large platoon moves out of the coverage area and needs to split into two smaller platoons to maintain the same level of QoS, it can sense other sub-channels. For example, with two divided platoons A and B, as long as there is a vacant sub-channel, the divided platoon B can use the vacant channel. By doing this, platoon B does not necessarily need to wait until platoon A moves out of the transmission range to avoid inter-platoon interference as shown in Fig. 6.

*C. Dynamic platooning*

From the previous discussions, we know that if the platoon is in the coverage area, network scheduled MAC protocols can be used and platoon size can be increased for higher road capacity. Once the platoon is out of the coverage area, network coordination is unavailable and autonomous MAC protocols need to be employed. However, with the same frequency band platoon size should be smaller. It means that when a platoon moves out of the coverage area, it may need to be split to maintain the same level of QoS and multiple platoons may need to be merged when moving into the coverage area to improve road capacity and traffic efficiency.

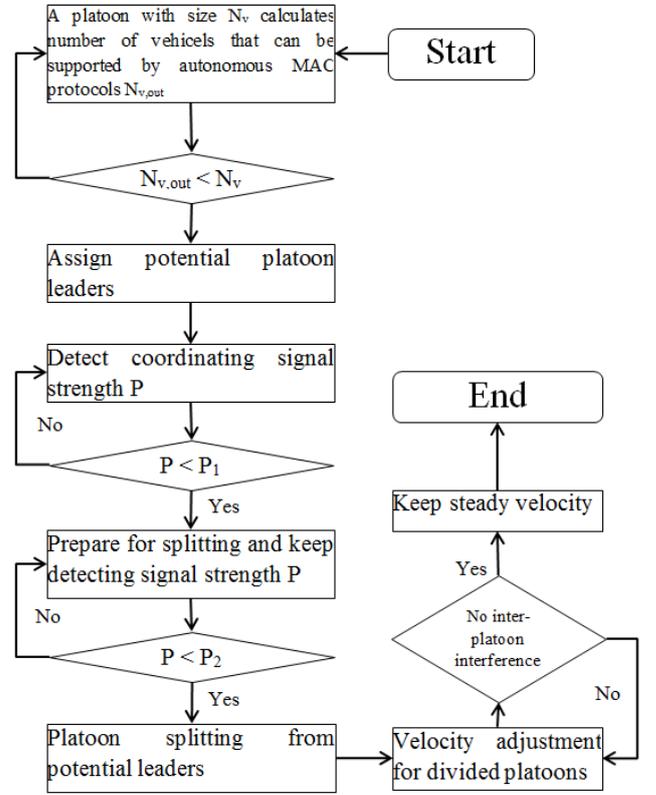

Figure 7 Platoon splitting flow chart

Platoon splitting happens when a platoon moves out of the coverage area. Since the splitting is triggered by moving out of coverage area, there should be a procedure to measure the coordinating signal strength from the BS. Once the signal strength is below certain threshold, it implies that the platoon is moving out the coverage area and splitting should happen. However, since the signal strength can be easily affected by fading and might be below the threshold even when the platoon is in the coverage area, the splitting should not happen immediately after the coordinating signal from BS gets weaker. In addition, the platoon also needs to prepare for the splitting. In this regard, we could have two thresholds $P_1$ and $P_2$. Once the signal strength is smaller than $P_1$, the platoon gets ready to split, e.g., start sensing the adjacent vacant sub-channels. If the signal strength is weaker than $P_2$, the platoon initiates the actual splitting procedure . Ideally, the split should happen in the middle of the platoon to minimize the coordination within the platoon and interruption to other road users. Since each platoon needs a leader, a platoon likely to split needs to assign a potential leader for the newly-formed platoon.

For single channel case, the divided platoons need to adjust their velocity to make sure they are not in the transmission range of each other to avoid inter-platoon interference. For example, one platoon may need to accelerate and the other one slows down and maintains a steady velocity once they are out of each other's transmission range. For sub-channel case, the platoon about the split needs to detect vacant sub-channel. Once there are available vacant sub-channels, the divided platoon can be assigned to those sub-channels to avoid inter-platoon interference. If no vacant sub-channel is available, the same procedure as single channel case can be used. The procedure for single channel is shown in Fig. 7.

Platoon merging operation can be easily coordinated by the eNB since it happens in the coverage area.

## IV. EVALUATION RESULTS

In this section, we evaluate the optimal platoon size and the evaluation parameters are listed in table 1 [5].

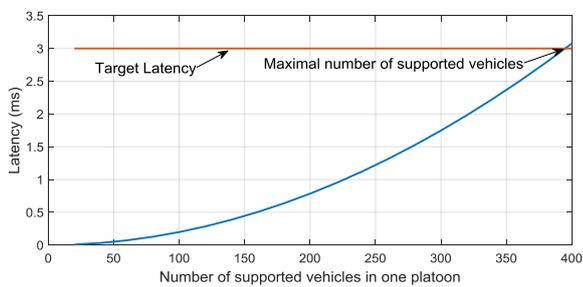
Figure 8 Latency (reservation based MAC)

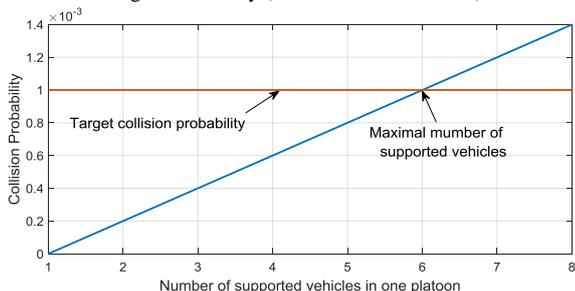
Figure 9 Reliability in terms of collision probability (slotted ALOHA)

Latency of reservation based MAC and collision probability of slotted ALOHA are illustrated in Fig. 8 and 9, respectively. Based on Fig. 8 and 9, the maximal numbers of supported vehicles for slotted ALOHA and reservation-based approach is 6 and 394, respectively. From this evaluation results, we can see that with the same service requirement and resource availability, the reservation-based MAC has a much larger MAC efficiency $\eta_{MAC}$ than the slotted ALOHA and therefore the number of supported vehicles, i.e., platoon size, is different. Thus a dynamic platooning mechanism is needed.

Table-1 Evaluation Parameters

| Parameter | Value |
|---|---|
| $L_{pkt}$ | 50 Bytes |
| $R_{gen}$ | 10 |
| $S_{mcs}$ | 2 |
| $B$ | 10MHz |
| $s$ | 1.5m |
| $d$ | 1m |
| $D$ | 50m |
| $v$ | 20m/s |
| Target reliability $P_{target}$ | 0.001 |
| Target latency $T_{target}$ | 3 ms |

## V. CONCLUSIONS

In this paper, we propose a dynamic HDVP mechanism based on an important observation that the vehicle platoon size is affected by resource availability and QoS requirements and should be adapted based on different MAC protocols employed, which means that the platoon needs to have the capability of splitting and merging to adapt the platoon size under different circumstances. With the proposed approach road capacity and traffic efficiency can be maximized without violating any QoS requirements, e.g., latency and reliability.

ACKNOWLEDGEMENTS

The European Commission funding is acknowledged for the phase II ONE5G project (ICT 760809), through which part of the research leading to this paper was conducted.